\documentclass[
  journal=jcisd8,
  manuscript=article,
  layout=traditional
]{achemso}
\usepackage{amsmath}
\usepackage{booktabs}
\usepackage{graphicx}
\usepackage{enumitem}
\usepackage{xcolor}
\usepackage{hyperref}

\usepackage{xr}
\setlist[itemize]{leftmargin=*,nosep}

\newcommand{\bcover}{BCover}

\title[\bcover{} Covalent Docking Scoring]{\bcover{}: An Electronic Structure-Based Scoring Suite for Reaction-Aware Covalent Docking}

\author{Emil Zak}
\email{emil@beit.tech}
\author{Michał Szczepanik}

\affiliation{BEIT sp. z o.o., Wadowicka 8A, 30-415 Krak\'ow, Poland}


\begin{document}

\begin{abstract}
Covalent virtual screening requires ranking compounds according to both noncovalent recognition and their ability to adopt a reaction-competent geometry with an appropriately reactive warhead. Here, we introduce \bcover{}, a reaction-aware scoring suite that combines pre-reactive docking with quantum-chemistry-derived ligand reactivity descriptors, the electrostatic properties of the protein pocket. Quantum-chemical descriptors are aggregated using a nonlinear tree-based scoring model. 

\bcover{} was evaluated retrospectively on the COValid benchmark, comprising nine targets and ten reactive sites, and compared with AutoDock, DOCK6, DOCKovalent, AlphaFold3 with Rosetta rescoring, and AlphaFold3 confidence-based ranking. \bcover{} achieved an average adjusted LogAUC of 31\% (57\% max), an average ROC-AUC of 0.88 (0.96 max), an average EF1 of 18 (33 max). Its average LogAUC exceeded those of the classical docking methods and AlphaFold3-Rosetta, although AlphaFold3-mPAE provided the strongest overall enrichment. At an average runtime of about 10~s per ligand, \bcover{} was approximately 25-fold faster than the evaluated AlphaFold3 workflows and achieved the highest average time-adjusted virtual-screening productivity index. Redocking experiments further showed that the method recovered near-native ligand conformations.

These results demonstrate that combining docking-derived geometry with ligand local electronic reactivity and pocket electrostatics provides an efficient and interpretable strategy for covalent ligand prioritization. \bcover{} is intended as a high-throughput screening method that complements more computationally demanding QM/MM and free-energy calculations during subsequent lead optimization.
\end{abstract}

\section{Introduction}
Covalent inhibitors are increasingly important in drug discovery because they can combine target recognition and selectivity with long residence times~\cite{powers2002proteases,backus2016proteome,flanagan2014reactivegroups}. 
Covalent ligand discovery requires simultaneous satisfaction of noncovalent recognition, i.e. molecular docking of the ligand molecule in the target, reaction-enabling geometry, and environment-modulated chemical reactivity~\cite{scarpino2018comparative,goullieux2023twostep,peng2025covdocker,palazzesi2019abinitio,hermann2020electrophilicity,yu2019kinetics,Shamir2026COValid}.

Pose reproduction alone is not sufficient for covalent docking. A near-native pose does not guarantee useful ranking in retrospective or prospective screening \cite{scarpino2018comparative,goullieux2023twostep,peng2025covdocker}. 
Prior covalent docking approaches primarily estimate noncovalent interaction energy and steric complementarity. Tethered or constrained covalent docking can improve geometry, but a geometrically plausible pose does not establish that the warhead is sufficiently electrophilic (reactive) or that the receptor environment favors charge transfer and bond formation~\cite{ouyang2013covalentdock,zhu2014covdock,scarpino2021widock,wu2023hcovdock,yu2023silcs,goullieux2025hybrid}.
Tools based on chemical reactivity of the ligand alone often combine global and local ligand reactivity indices. Those that use global ligand descriptors, such as ionization energy, electron affinity, chemical hardness, and global electrophilicity, can characterize intrinsic molecular reactivity but do not identify the reactive atom or account for the particular protein pocket. Conversely, local geometry-dependent electronic reactivity terms depend on the selected pose but generally omit the electronic polarization imposed by the receptor. Finally, tools based on ligand reactivity alone lack appropriate pre-reactive complex generation penalty contributions.

The practical gap addressed in this work is a fast, automated scoring tool that keeps the interpretability of physics-inspired scoring and provides a sufficiently high ligand enrichment power in covalent inhibitors screening.
By forming descriptors that combine local ligand response functions with a spatially resolved receptor potential and pre-complex pose optimization, our model can differentiate candidate poses and ligands that are similar according to docking energy or global electrophilicity but differ in (i) localization of electrophilic response at the warhead, (ii) electrostatic activation or deactivation by the pocket, (iii) reaction-center geometry, and (iv) conformational preorganization at the noncovalent docking stage.

In this work, we present \bcover{}, a fast, automated, reaction-aware covalent docking and scoring suite that improves retrospective covalent virtual-screening performance by combining pre-reactive docking with pose scoring based on quantum-chemical electronic structure calculations.
Our contributions are as follows. We introduce a new scoring formula based on reactivity descriptors that combine Fukui-function-based reactivity and quantum-chemically calculated electrostatic information about the pocket encoded in electrostatic potential. Our method enables pocket detection, reactive residue detection, warhead detection, pocket-ligand interaction profiling, pre-reactive molecular docking with quantum-inspired solver, and pose scoring. We carried out retrospective benchmarking on the COValid benchmark set~\cite{Shamir2026COValid}, which contains nine protein targets and ten target-residue sites, and compared \bcover{} with widely used tools for covalent virtual screening, including AutoDock~\cite{morris2009AutoDock4}, DOCK6~\cite{Allen2015DOCK6}, DOCKovalent~\cite{London2014DOCKovalent}, CarsiDock-Cov~\cite{Shen2025CarsiDockCov}, and AlphaFold3~\cite{abramson2024alphafold3}. All the benchmarked tools are open source. 

\bcover{} demonstrated ligand-ranking capability that outperformed available open-source and commercial physics-based scoring approaches, while matching the accuracy of AlphaFold3 with Rosetta rescoring~\cite{abramson2024alphafold3,davis2009rosettaligand}, as measured by adjusted LogAUC, BEDROC, enrichment factor at 1\% (EF1), and ROC-AUC. Although \bcover{} achieved lower ranking metrics than AlphaFold3 using the predicted aligned error matrix metric (AF3-mPAE)~\cite{Shamir2026COValid}, it was approximately 25-fold faster on average. \bcover{} therefore operates in a runtime regime comparable to lightweight tools such as AutoDock and DOCK6, while offering an efficient multi-CPU parallelization for pre-reactive docking stage based on simulated annealing of a cost function based on classical MMFF94 force field and LM5 entropy model and multi-GPU parallelization of selected modes in the quantum-computational stage of our workflow, making it suitable for high-throughput covalent virtual screening.

\section{Methods}

\subsection{Overview of the \bcover{} Workflow}
The main contribution of this work is discussed in sec.~\ref{sec:scoring}, where we present details of our quantum pose scoring function. In this section, we briefly discuss key stages of our design, which overall represents a beginning-to-end pipeline, which we refer to as \bcover{}.

\begin{figure}[htbp]
    \centering
    \includegraphics[width=\linewidth]{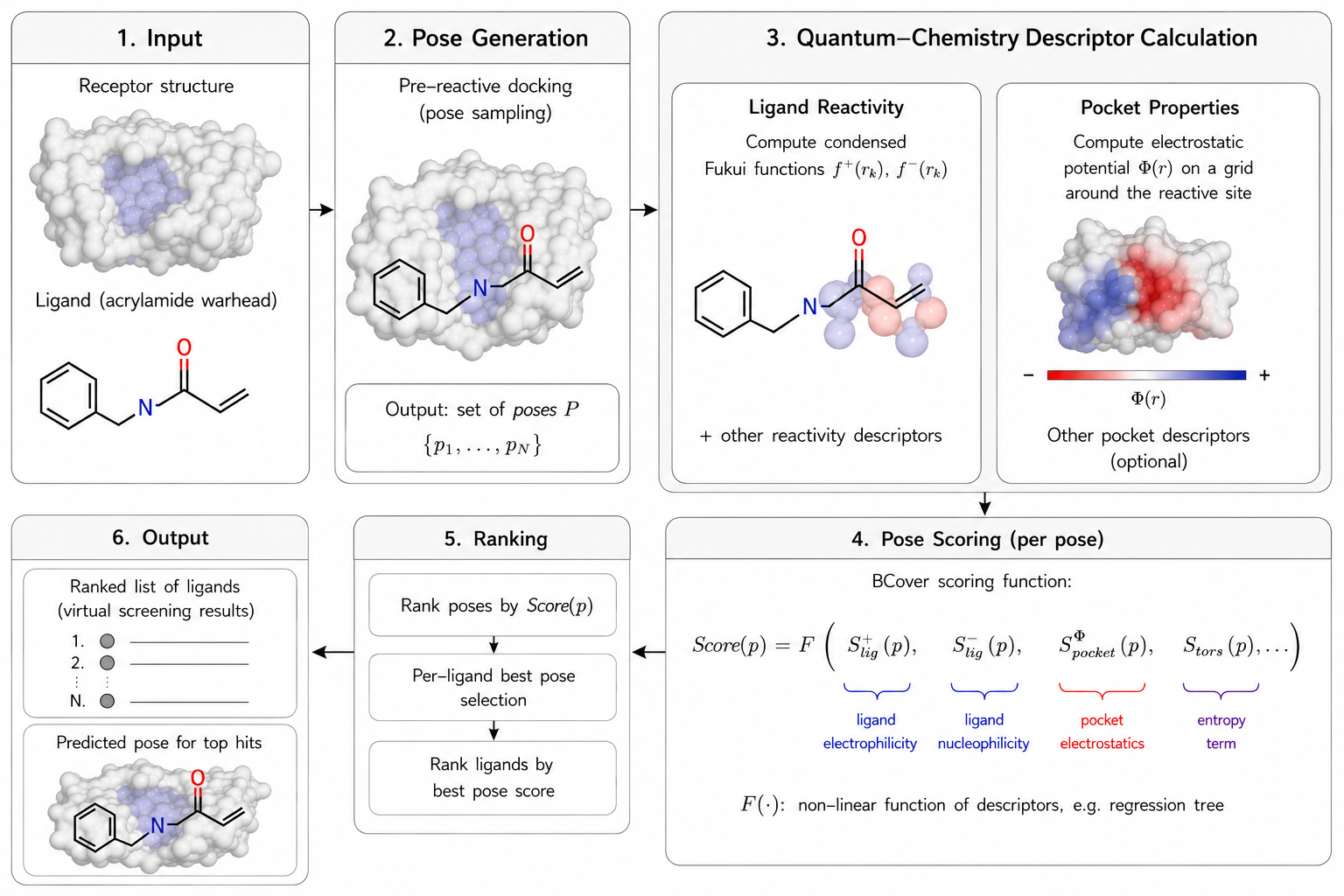}
    \caption{Schematic of the BCover pipeline.}
    \label{fig:model}
\end{figure}

\bcover{} is a computer code implementing an automated pipeline for reaction-aware virtual screening of covalent inhibitors. Starting from a protein structure and a ligand library, the  code identifies candidate reactive sites and lignd warheads, generates pre-reactive ligand poses, evaluates their geometric and quantum-chemical compatibility with covalent bond formation, and returns a ranked list of ligands together with interpretable score components (Fig.~\ref{fig:model}).

\paragraph{Input.}
The required input consists of (i) a protein structure in PDB format, (ii) a ligand database provided as SDF, MOL2 or SMILES, and optionally (iii) user-defined binding-site identifier (reactive residue ID) or reaction-specific geometric constrains for residue-warhead near-attack orientation. If no constraints are supplied, all subsequent stages are performed automatically.

\paragraph{Preprocessing.}
Potential ligand-binding pockets are identified using FPocket~\cite{leGuilloux2009fpocket}. Alternatively, the user may directly specify the binding pocket by residue identifiers.
For every detected pocket, \bcover{} identifies aminoacid residues capable of forming covalent bonds with electrophilic ligands. The current implementation supports the most common nucleophilic residues exploited in covalent drug discovery, including cysteine, serine, lysine, tyrosine, histidine and threonine. Each residue is represented by one or more reactive atoms corresponding to the expected nucleophilic center (e.g. S$_\gamma$ for cysteine, O$_\gamma$ for serine, N$_\zeta$ for lysine and O$_\eta$ for tyrosine).

The docking simulation box is subsequently generated automatically. Its center can be defined by the reactive atom, the geometric center of the detected pocket, or the center of a co-crystallized ligand when available. The box size is determined from the ligand library according to $ L = \alpha \max_i \left(
\max_{a,b\in i} \|\mathbf r_a-\mathbf r_b\|\right)$, where $L$ denotes the cubic box length, $\alpha=2$ by default is a scaling factor, and the inner maximum corresponds to the largest intramolecular atomic distance for ligand $i$. This procedure ensures that all candidate ligands can be sampled without artificial steric truncation.

Ligands are simultaneously analyzed for the presence of reactive warheads using SMARTS-based substructure recognition. The current implementation recognizes several clinically relevant electrophilic warhead classes including nitriles, acrylamides and related Michael acceptors.

\paragraph{Pre-reactive docking.}
Ligand conformations are generated using the BDocker docking engine based on simulated annealing~\cite{BEIT2026BDocker}. During docking a cost function is minimized, which combines ligands' internal potential energy and non-covalent ligand-pocket interaction terms. During the optimization, ligand internal energy is described by a MMFF94 force field including bond stretching, angle bending, torsional, stretch-bend coupling, while the ligand-pocket interaction model contains electrostatic and van der Waals interactions together with an implicit desolvation and hydrogen-bonding contributions. 

Unlike conventional non-covalent docking, \bcover{} explicitly biases sampling toward reaction-competent geometries through a physically motivated confining potential similar in spirit to previously proposed two-step covalent docking approaches~\cite{ouyang2013covalentdock,zhu2014covdock,goullieux2023twostep}. The restraint acts only on the reactive center and guides the electrophilic warhead toward the nucleophilic residue while preserving full conformational flexibility of the remaining ligand. The modification to the force field includes additional harmonic potential confining terms,  which ensure residue-warhead orientation in a near-attack conformation (NAC) geometry, derived from quantum-chemical calculations. These calculations are carried per chemical reaction. For the case of Michael addition benchmarked in this work we used DFT with $\omega$B97X-D3BJ functional and the aug-cc-pvdz basis set, to obtain the NAC geometry used to parameterize the geometric restraints employed during docking.

For each ligand, the code returns an ensemble of low-energy pre-reactive poses. Depending on the application, either the lowest-energy pose or a Boltzmann-weighted ensemble may be propagated to the scoring stage. A schematic of the adopted model is presented in Fig.~\ref{fig:model}.

\paragraph{Pose scoring.}
Each pose is evaluated using a multi-component scoring function, which accounts for electronic reactivity, pose-penalty from the pre-reactive stage and entropic contributions. 
Electronic reactivity is characterized using quantum-chemistry-derived descriptors including Fukui functions and related electrophilicity measures~\cite{lonsdale2017warhead,palazzesi2019abinitio,hermann2020electrophilicity}. These descriptors are described in detail in sec.~\ref{sec:scoring}. The final score additionally incorporates the LM5 entropy model~\cite{Chan2021} to estimate conformational entropy losses associated with ligand binding.

\paragraph{Output.}
For every ligand, \bcover{} returns ranked poses, overall scores, decomposition into individual score components, and execution metadata including runtime and statistics. These outputs provide both high-throughput virtual-screening capability and physically interpretable information suitable for a follow-up analysis.

\subsection{Reaction-aware quantum-chemical pose scoring}
\label{sec:scoring}

Ligand poses are scored with a function that can be expressed as follows:
\begin{equation}\label{eq:score}S_{\mathrm{BCover}}=f(
 S_{\mathrm{dock}},
 S_{\mathrm{int}},
 S_{\mathrm{react}},
 S_{\mathrm{ent}}).
\end{equation}
where $S_{\mathrm{dock}}$ captures noncovalent recognition and pose penalty, $S_{\mathrm{int}}$ describes electrophile-nucleophile compatibility using Fukui-function-derived descriptors \cite{lonsdale2017warhead,palazzesi2019abinitio,hermann2020electrophilicity} as well as molecular-potential contributions \cite{lepsik2013sqm,smith2013rangeseparated,smith2015thiol},$S_{\mathrm{react}}$ captures intrinsic pose reactivity using quantum-chemistry-derived descriptors \cite{lepsik2013sqm,smith2013rangeseparated,smith2015thiol} and $S_{\mathrm{ent}}$: accounts for entropy-related penalties using the LM5 entropy model~\cite{Chan2021}.

\begin{figure}[htbp]
    \centering
    \includegraphics[width=0.55\linewidth]{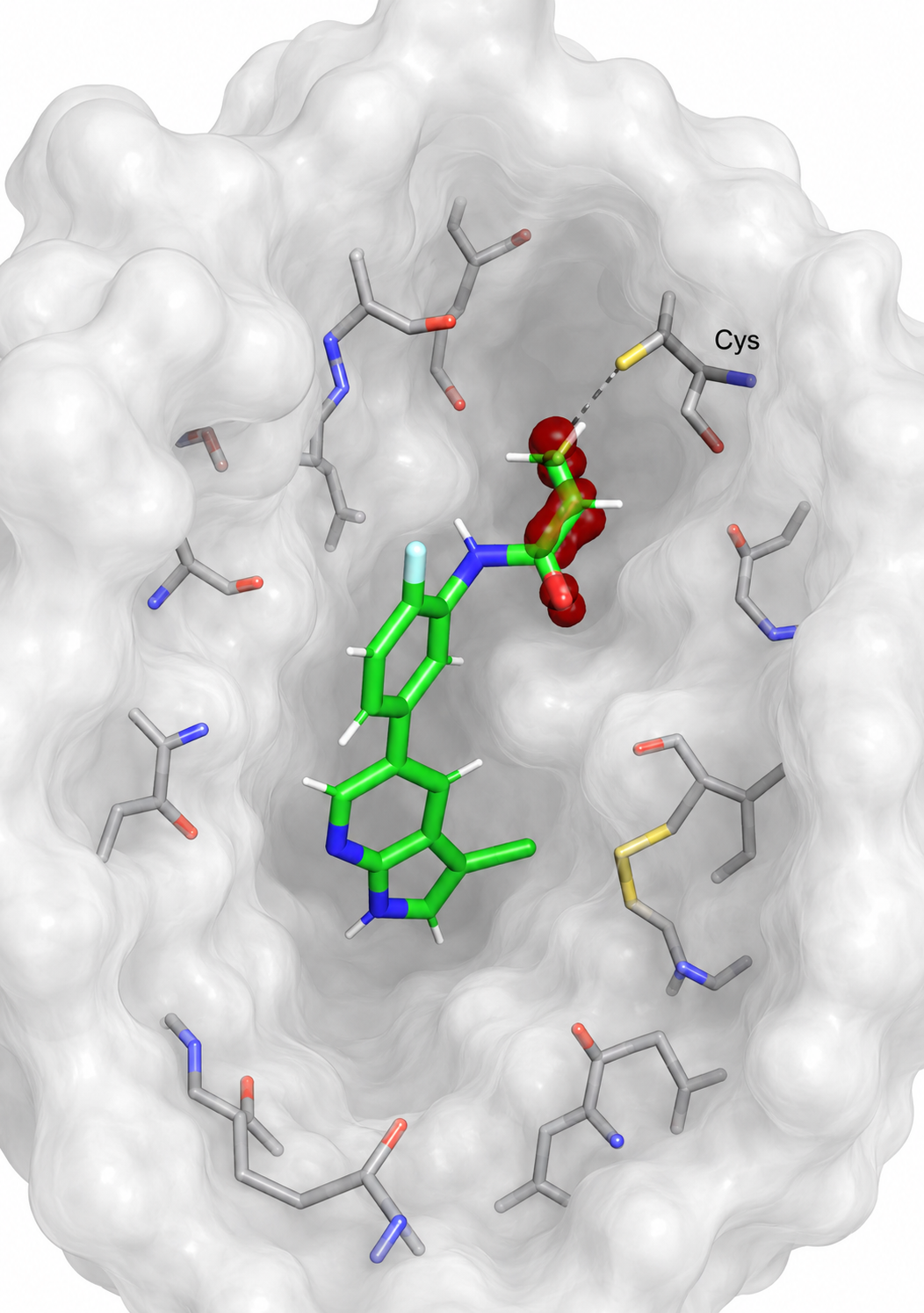}
    \caption{
    A representation of the target pocket with covalently bound ligand. Shown is also isosurface representing the Fukui function near the warhead region. }
    \label{fig:fukui_pocket}
\end{figure}

The set of calculated reactivity descriptors $S_{react}$ consists of ionization energy $I$, electron affinity $A$, chemical potential $\mu \approx \frac{1}{2}(I+A)$, hardness $\eta \approx \frac{1}{2}(I-A)$ and global electrophilic index $GEI = \frac{\mu^2}{2\eta}$. 
Global descriptors such as $GEI$ have been shown to correlate with experimental data on covalent inhibitors on certain datasets \cite{rowan_scientific_2026,hermann2020electrophilicity}. However, these quantities lack any information about the target protein or local reactivity of the ligand. We solve this by introducing another set of descriptors $S_{int}$ quantifying the interaction between the ligand and the pocket. The information about local reactivity of the ligand is included using Fukui function, which can be expressed \cite{ph15091112}
\begin{align}\label{eq:fukui_volumetric}
    f^{(+)}(\boldsymbol{r}) \approx \rho_{N+1}(\boldsymbol{r})-\rho_N(\boldsymbol{r}) && f^{(-)}(\boldsymbol{r}) \approx \rho_{N}(\boldsymbol{r})-\rho_{N-1}(\boldsymbol{r})
\end{align}
where $\rho_N(\boldsymbol{r})$ denotes electron density corresponding to $N$ electron system. Function $f^{(+)}(\boldsymbol{r})$ describes reactivity towards nucleophiles, while $f^{(-)}(\boldsymbol{r})$ describes reactivity towards electrophiles. One can also assign a Fukui index to each atom
using
\begin{align}\label{eq:fukui_atomic}
    f^{+}_k(\boldsymbol{r}) = q_k(N) - q_k(N+1) && f_k^{-}(\boldsymbol{r}) = q_k(N-1) - q_k(N)
\end{align}
where $q_k(N)$ is an atomic charge of $k$-th atom obtained from methods such as Mulliken or Löwdin population analysis. To incorporate information about the target, the calculated descriptors also depend on the external potential experienced by the ligand within the binding pocket, which within DFT framework can be expressed as\cite{Morell2004}
\begin{equation}
    V_{poc}(\boldsymbol{r}) = -\sum_{k=1}^{N_{poc}}\frac{Z_k}{|\boldsymbol{r}-\boldsymbol{R}_k|}+\int d\boldsymbol{r}'\frac{\rho_{poc}(\boldsymbol{r}')}{|\boldsymbol{r}-\boldsymbol{r'}|}+v^{XC}(\boldsymbol{r})
\end{equation}
where $Z_k$ and $R_A$ denote the atomic number and the position of $k$-nuclei, $N_{poc}$ is the total number of atoms in the pocket, $\rho_{poc}(\boldsymbol{r'})$ is electronic density inside the pocket and $v^{XC}(\boldsymbol{r})$ is the exchange correlation potential. Based on eqs~\ref{eq:fukui_volumetric},\ref{eq:fukui_atomic} we introduce the following derived descriptors:
\begin{equation}
    F = \frac{\mu}{\eta}\sum_{k\in Warhead} f^{(+)}_k
\end{equation}
\begin{equation}\label{eq:E_f}
    E_f = \int d\boldsymbol{r} f^{(+)}(\boldsymbol{r})V_{poc}(\boldsymbol{r})
\end{equation}
\begin{align}
    C_{cov}^{(2)} = \frac{\mu}{\eta}E_f  && C_{cov}^{(3)} = \frac{1}{2\eta}E_F^2
\end{align}
Those quantities are inspired by an approximate expression for covalent contribution to the binding energy \cite{Morell2004} 
\begin{equation}\label{eq:E_cov}
    E_{cov} = \frac{
    \left[\mu_{lig}-\mu_{poc}+\int d\boldsymbol{r}f_{lig}^{(+)}(\boldsymbol{r})V_{poc}(\boldsymbol{r})-\int d\boldsymbol{r}f_{poc}^{(-)}(\boldsymbol{r}) V_{lig}(\boldsymbol{r})\right]^2}{2(\eta_{lig}+\eta_{poc})}
\end{equation}
where $V_{lig}(\boldsymbol{r})$ and $V_{poc}(\boldsymbol{r})$ are molecular electrostatic potentials generated  by the ligand and the receptor, respectively. Based on eq.~\ref{eq:E_cov} we define a related quantity
\begin{equation}\label{eq:C_cov}
    C_{cov} = \frac{\left[\mu_{lig} + \int d\boldsymbol{r}f^{(+)}_{lig}(\boldsymbol{r})V_{poc}(\boldsymbol{r})\right]^2}{2\eta_{lig}} = GEI+C_{cov}^{(1)}+C_{cov}^{(2)}
\end{equation}
Descriptors $C_{cov}^{(1)}$ and $C_{cov}^{(2)}$ can therefore be interpreted as corrections to GEI, which include local reactivity information via $f^+{(\boldsymbol{r})}$ and receptor information via $V_{poc}(\boldsymbol{r})$. An example isosurface plot of the Fukui function for a representative ligand from the Covalid dataset is displayed in Fig.~\ref{fig:fukui_pocket}.

To reduce the runtime of the calculation one can use the following approximation to the interaction term $E_f$
\begin{equation}\label{eq:E_f_atomic}
    E_f \approx E_f^{(atomic)} = \sum_{k=1}^{N_{lig}}f^{(+)}_{k}V_{poc}(\boldsymbol{r}_k)
\end{equation}
where $\boldsymbol{r}_k$ denotes the position of $k$-th atom and $N_{lig}$ is the total number of atoms in the ligand. Our initial tests showed that such an approach does not results in significant loss of accuracy on considered datasets, as seen in Table \ref{supp-tab:method_comparison_performance} in Supporting Information. 

Our final scoring function given in eq.~\ref{eq:score} was obtained by training a model, which takes calculated molecular descriptors as input and outputs the score $S_{BCover}\in [0,1]$, which can be interpreted as a predicted probability that a given ligand is an active binder. We utilized the extreme gradient boosting (XGB)~\cite{Chen_2016} model, which is a variant of tree-based classification method. Binary decision trees are non-parametric supervised learning method that recursively partition the data by selecting, at each node, the split that minimizes a cost function measuring the impurity of the resulting child nodes~\cite{Loh2011}. While simple decision trees have several advantages, such as interpretability and the ability to model complex nonlinear relationships, they also suffer from several limitations, including poor generalization and a tendency to overfit. To address these shortcomings, several ensemble methods have been proposed, including the Random Forest algorithm\cite{breiman2001} and gradient boosting techniques\cite{friedman2001}. In particular, gradient boosting works by iteratively building multiple trees, such that every new tree is trained to correct the mistakes of previews ones. XGB is a highly optimized implementation of the gradient boosting algorithm, known for its speed, efficiency and scalability~\cite{Chen2016XGBoost}. 

The final model used in \bcover{} was trained on the COValid dataset. For each target, the set of ligands was partitioned into training and test data with equal sizes and the same active to decoy ratio. As described in Supporting information, we found the optimal set of input descriptors for the model to be $\overline{X} =\{C_{cov}^{(1)},\mu,F,GEI,\eta,S_{ent}\}$,
where $E_{binding}$ is the binding affinity estimated during non-covalent docking stage.

\subsubsection{Quantum chemical calculations}
We implement two methods of calculating quantum-mechanical descriptors described in the previous section: density functional theory (DFT) and extended tight binding (xTB). For DFT we rely on the PySCF~\cite{Sun2017} implementation, while xTB calculations were performed using the xtb program package developed by the Grimme group~\cite{Bannwarth2020}. Evaluating descriptors such as the chemical potential or the Fukui functions requires three separate electronic structure calculations for each ligand, corresponding to the neutral, cationic, and anionic states. To reduce the computational cost of the DFT calculations, we also implemented an additional execution mode based on frozen orbital approximation, which we denote DFT(frozen). This allows one to express all descriptors in terms of properties related only to the neutral species:
\begin{align}
    I \approx  -\epsilon_{HOMO} && A \approx  -\epsilon_{LUMO} && f^{+}(\boldsymbol{r}) \approx  |\psi_{LUMO}|^2 && f^{(-)}(\boldsymbol{r}) \approx  |\psi_{HOMO}|^2
\end{align}
where $\epsilon_{HOMO}$, $\epsilon_{LUMO}$ denote orbital energies on Highest Occupied Molecular Orbital (HOMO) and Lowest Unoccupied Molecular Orbital (LUMO) respectively and $\psi_{HOMO}$, $\psi_{LUMO}$ are the corresponding wavefunctions.
Table~\ref{supp-tab:method_comparison_r_squared} in Supporting Information shows the comparison between DFT, xTB and DFT with frozen orbital approximation. Due to the favorable computational cost, we focus on the xTB implementation in the rest of this paper.

\subsection{Benchmark Dataset}
\label{subsec:benchmark_dataset}
We used COValid as the primary retrospective benchmark for evaluating covalent virtual-screening performance~\cite{Shamir2026COValid}. COValid was selected because it was designed specifically for covalent enrichment analysis rather than only pose reproduction, and therefore provides a suitable test of whether a scoring function can prioritize active covalent ligands over property-matched decoys. The benchmark was interpreted in the context of prior covalent-docking benchmark studies, which report that structure curation is important as well as pose and reactivity evaluation beyond simple redocking success~\cite{scarpino2018comparative,goullieux2023twostep,peng2025covdocker,singh2025landscape}.

The COValid set contains nine protein targets and ten annotated nucleophilic attachment sites. The benchmark focuses on cysteine-targeting acrylamide ligands, dominated by cysteine-directed covalent inhibitors and providing a chemically controlled setting for comparing enrichment performance. For each active ligand, COValid includes a set of property-matched decoy protomers generated by removing and reinstalling an acrylamide electrophile on topologically dissimilar noncovalent analogues. This design reduces trivial property-based separation between actives and decoys while retaining chemically plausible covalent warheads. In total, the benchmark contains 874 active protomers and 37,919 decoy protomers distributed across the ten target sites.

Active labels in COValid are derived from experimental activity annotations curated from BindingDB and ChEMBL~\cite{Gilson2016BindingDB,Dallago2024ChEMBL34}, with filtering steps intended to enrich for compounds acting through a covalent cysteine-targeting mechanism~\cite{Shamir2026COValid}.  Because decoys are not necessarily experimentally confirmed inactives, we treat COValid as a retrospective enrichment benchmark rather than a direct estimate of absolute biochemical classification accuracy. Accordingly, our analysis includes comparing relative ranking metrics, including adjusted LogAUC, BEDROC, EF1, ROC-AUC, and time-adjusted screening productivity, which we define further on.

For each target site, we retained the COValid annotations of the reactive residue, target identity, active and decoy protomer counts, and associated structural references. Where pose-level analysis was possible, predicted poses were compared to the corresponding experimental covalent reference structures using pocket-aligned ligand RMSD. 

An example output including target name, active/decoy annotations, protomer ID, descriptors and final BCover score is reported in the Supporting Information. This per-target reporting is important because covalent virtual-screening performance can vary substantially with target family, reactive-site location, ligand chemotype distribution, and the quality of active/decoy annotations.

\subsection{Comparator Tools}
\label{subsec:comparators}
\bcover{} was benchmarked against representative state-of-the-art methods for covalent virtual screening, including AutoDock4~\cite{morris2009AutoDock4}, DOCK6~\cite{Allen2015DOCK6}, DOCKovalent~\cite{London2014DOCKovalent}, CarsiDock-Cov~\cite{Shen2025CarsiDockCov}, AlphaFold3 with both the predicted aligned error (AF3-mPAE) ranking and Rosetta rescoring (AF3--Rosetta)~\cite{abramson2024alphafold3,davis2009rosettaligand}. These methods span classical force-field docking plus covalent docking heuristics, deep-learning-based structure prediction, and hybrid structure-prediction/rescoring workflows.
AutoDock4, DOCK6, and DOCKovalent are physics-based docking tools that generate covalent or covalent-like ligand poses in a mostly fixed receptor structure and rank them with empirical or grid-based interaction scores. AutoDock treats covalent docking through constrained ligand placement and torsional sampling, DOCK6 uses an attach-and-grow strategy to construct the ligand from the reactive site, and DOCKovalent samples ligand conformations under idealized covalent bond-length and bond-angle constraints. AF3-Rosetta separates structure prediction from energetic rescoring: AlphaFold3 first predicts the covalent protein-ligand complex, after which Rosetta minimization and scoring are used to rank the resulting structures. AF3-mPAE instead uses AlphaFold3's predicted aligned error metric directly as a confidence-based ranking score, rather than an explicit physical interaction energy. A tool available through Rowan~\cite{rowan_scientific_2026} is treated here as a fast quantum/reactivity-oriented baseline that estimates covalent ligand suitability from electronic structure features rather than from full protein-ligand pose optimization. 

Whenever possible, all methods were evaluated using identical input data, including the same receptor structures, ligand libraries, protonation states, stereochemistry assignments, active/decoy annotations, and performance metrics. 

\begin{table}[ht]
\centering
\caption{Comparison of tools used in the Covalid benchmark and this work.}
\label{tab:comparators}
\begin{tabular}{lcccc}
\toprule
Method & Covalent-aware & Physics-based & ML-based & Open source \\
\midrule
AutoDock4 & No & Yes & No & Yes \\
DOCK6 & Partial & Yes & No & Academic \\
DOCKovalent & Yes & Yes & No & Yes \\
CarsiDock-Cov & Yes & Partial & Yes & Yes \\
AF3-mPAE & Partial & No & Yes & No \\
AF3-Rosetta & Partial & Partial & Yes & Partial \\
\bcover{} & Yes & Yes & Partial & No \\
\bottomrule
\end{tabular}
\end{table}

\section{Key metrics}
The receiver operating characteristic (ROC) curve is obtained by plotting the true positive rate (TPR) as a function of the false positive rate (FPR) over all possible decision thresholds\cite{Fawcett2006}. The area under the ROC curve (ROC-AUC) is defined as\cite{Hanley1982}
\begin{equation}
    \text{ROC-AUC} = \int_{0}^{1}\,dx TPR(x)
\end{equation}
where $x=\mathrm{FPR}$. To place greater emphasis on early enrichment (i.e., small FPR values), the ROC curve can instead be parameterized by log(FPR)\cite{Truchon2007}. The resulting metric is
\begin{equation}
    \mathrm{logAUC}(a)
    = \int_{a}^{1}d(\log x) TPR(x)
    = \int_{a}^{1}dx \frac{TPR(x)}{x}
\end{equation}
where $0<a<1$ is a small positive constant introduced to avoid the singularity at $x=0$. Throughout this work, we use $a=0.001$, following common practice~\cite{Shamir2026COValid,Mysinger2010,Mysinger2012,Bender2021}. One can also define an adjusted $\mathrm{logAUC(a)}$, which we simply denote by $\mathrm{logAUC}$, by subtracting the expected value for a random classifier,
\begin{equation}
    \mathrm{logAUC} = \mathrm{logAUC}(a) -\mathrm{logAUC}_{\mathrm{rand}}(a).
\end{equation}
Note that this metric still implicitly depend on the free parameter $a$.

A similar idea of assigning higher importance to early recognition is used in the definition of the Robust Initial Enhancement (RIE) metric~\cite{Sheridan2001}, which introduces an explicit exponential weighting of the ranked list:
\begin{equation}
    \mathrm{RIE}(\alpha)
    =
    \frac{\alpha}{N_A}
    \frac{\sum_{i=1}^{N_A} e^{-\alpha \frac{r_i}{N}}}
    {1-e^{-\alpha}},
\end{equation}
where $r_i$ denotes the rank of the $i$-th active ligand, $N_A$ is the total number of active ligands, $N$ is the total number of ranked compounds, and $\alpha$ is a parameter controlling the strength of the weighting towards the top of the ranking list. A disadvantage of RIE is that its minimum and maximum achievable values depend on the particular dataset and the number of active compounds. Therefore, the Boltzmann-enhanced discrimination of the receiver operating characteristic (BEDROC) metric is commonly used, which corresponds to a normalized version of RIE~\cite{Truchon2007}:
\begin{equation}
    \mathrm{BEDROC}(\alpha)=\frac{\mathrm{RIE}(\alpha)-\mathrm{RIE}_{\min}(\alpha)}{\mathrm{RIE}_{\max}(\alpha)-\mathrm{RIE}_{\min}(\alpha)}.
\end{equation}
Throughout this work, we use the values of $\alpha=80.5$.

Another important metric is the enrichment factor (EF), which quantifies the improvement in the fraction of active compounds retrieved at the top of a ranked list compared to a random classifier. It is defined as\cite{Truchon2007}
\begin{equation}
    \mathrm{EF}(\eta)=\frac{N_{\mathrm{active}}(\eta)/(\eta N)}
    {N_{\mathrm{active}}/N}=
    \frac{N_{\mathrm{active}}(\eta)}
    {\eta N_{\mathrm{active}}},
\end{equation}
where $\eta$ denotes the fraction of compounds considered at the top of the ranking list, $N$ is the total number of compounds, $N_{\mathrm{active}}(\eta)$ is the number of active compounds within the top $\eta$ fraction of the ranked list, and $N_{\mathrm{active}}$ is the total number of active compounds in the dataset. An EF$(\eta)$ value greater than one indicates enrichment relative to a random classifier. In this work, we mostly focus on cases of $\eta=0.01$ and $\eta=0.05$ and we denote the corresponding metrics as EF1 and EF5.

\section{Results}
\label{sec:results}

Below we report benchmark calculations results carried out on the COValid dataset. For \bcover{}, a single pre-reactive pose was retained for each ligand after docking and propagated to the scoring stage. During pose generation, the reactive geometry was guided by a harmonic confining potential acting on the electrophile-nucleophile pair. For acrylamide warheads reacting with cysteine residues, the equilibrium sulfur-C$_\beta$ distance was set to 2.3~\AA{} with a force constant of $k_r = 100$ kcal\,mol$^{-1}$\,\AA$^{-2}$. An additional angular restraint favored a C$_\beta$-S$_\gamma$-C$_{\mathrm{Cys}}$ angle of $90^\circ$, corresponding to the preferred near-attack geometry for Michael addition. These parameters were derived from density-functional calculations of representative model reactions and were kept fixed throughout all benchmark calculations.  
All ligand-specific descriptors were computed using the GFN2-xTB method~\cite{Bannwarth2019}, as implemented in the xTB package~\cite{Bannwarth2020}. The approximation given by Eq.~\ref{eq:E_f_atomic} was employed throughout. Atomic Fukui indices were calculated using Mulliken population analysis. The external potential $V_{\mathrm{pot}}(\mathbf{r}_k)$ evaluated at the positions $\mathbf{r}_k$ of the ligand atoms, was computed using DFT with the B3LYP exchange-correlation functional and the 6-31G* basis set, employing the density fitting approximation. The integration required to evaluate the electron contribution to external potential was performed using \texttt{df.incore.aux\_e2} function implemented in PySCF. All parameters unspecified it the text were kept at default values.

\subsection{Retrospective Screening Performance}
\begin{figure}[htbp]
    \centering
    \includegraphics[width=\linewidth]{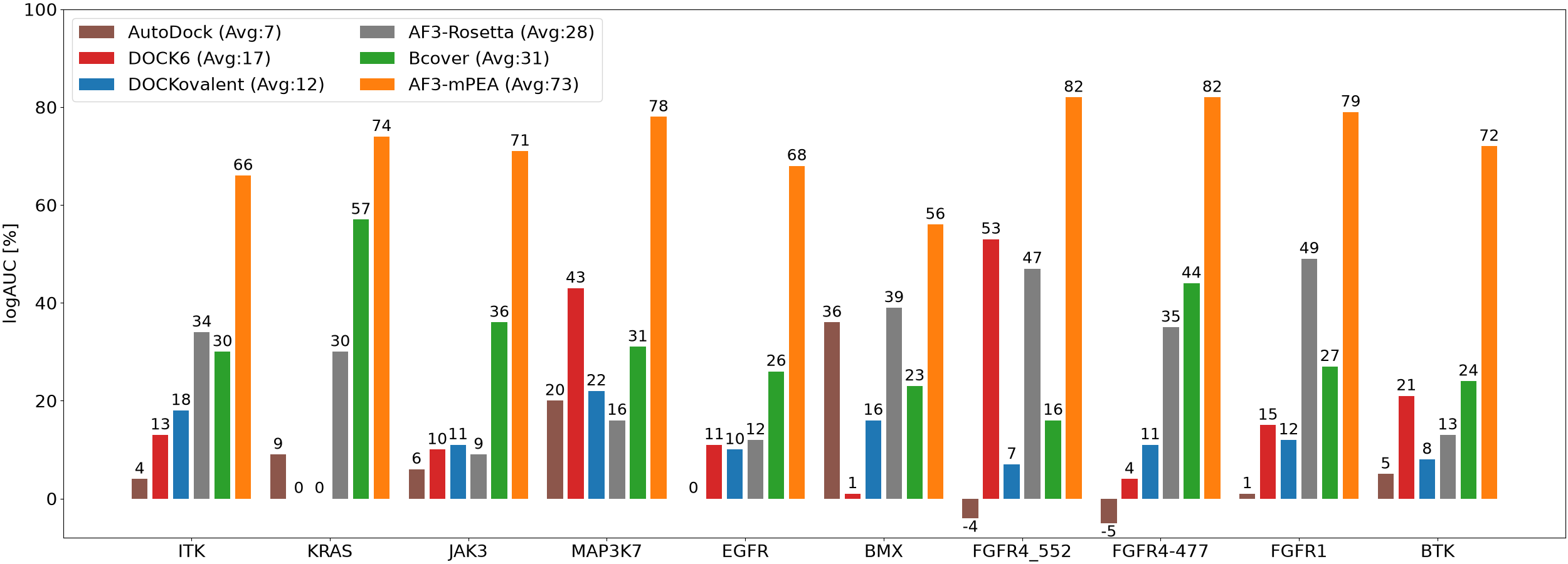}
    \caption{Per-target retrospective virtual-screening performance on COValid.
    Bars show adjusted logAUC values for each target site and method.
    Average adjusted logAUC values across targets are shown in the legend.}
    \label{fig:logauc_per_target}
\end{figure}

\begin{figure}[htbp]
    \centering
    \includegraphics[width=0.55\linewidth]{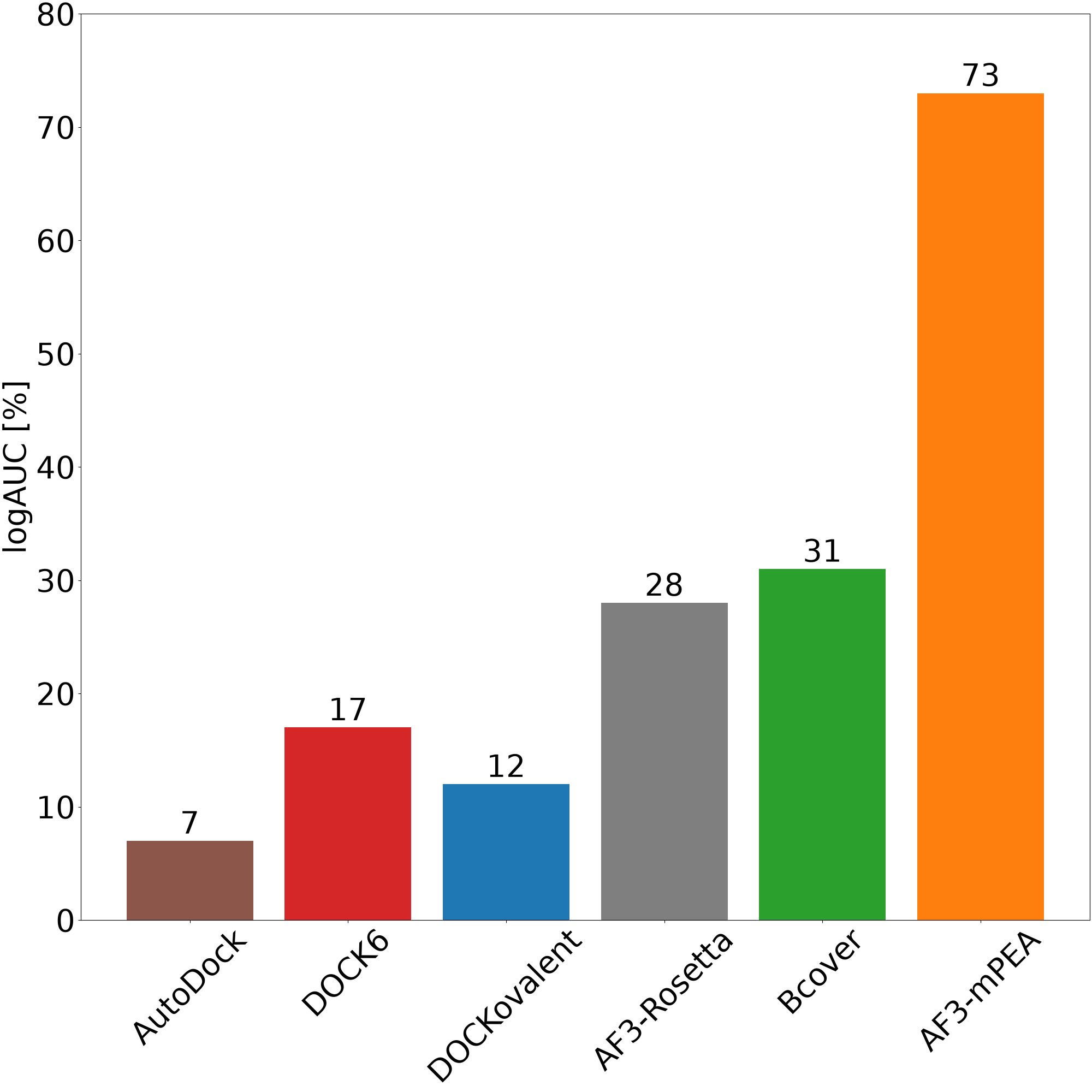}
    \caption{    Average adjusted logAUC across COValid target sites.
    Bars compare AutoDock, DOCK6, DOCKovalent, AF3-Rosetta, \bcover{}, and AF3-mPAE.
    }
    \label{fig:logauc_average}
\end{figure}

\begin{figure}[htbp]
    \centering
    \includegraphics[width=\linewidth]{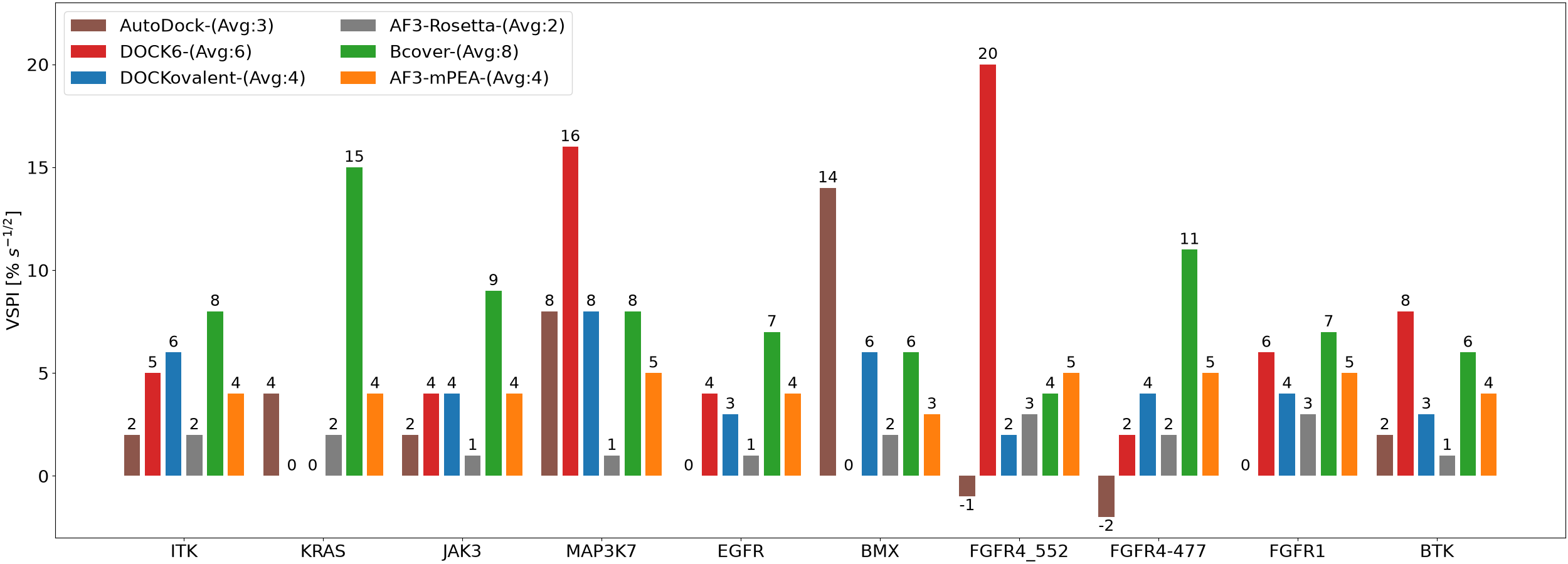}
    \caption{Per-target virtual-screening productivity index (VSPI) on COValid.
    Bars show VSPI values for each target site and method.
    Average VSPI values across targets are shown in the legend.}
    \label{fig:vspi_all}
\end{figure}

\begin{figure}[htbp]
    \centering
    \includegraphics[width=0.55\linewidth]{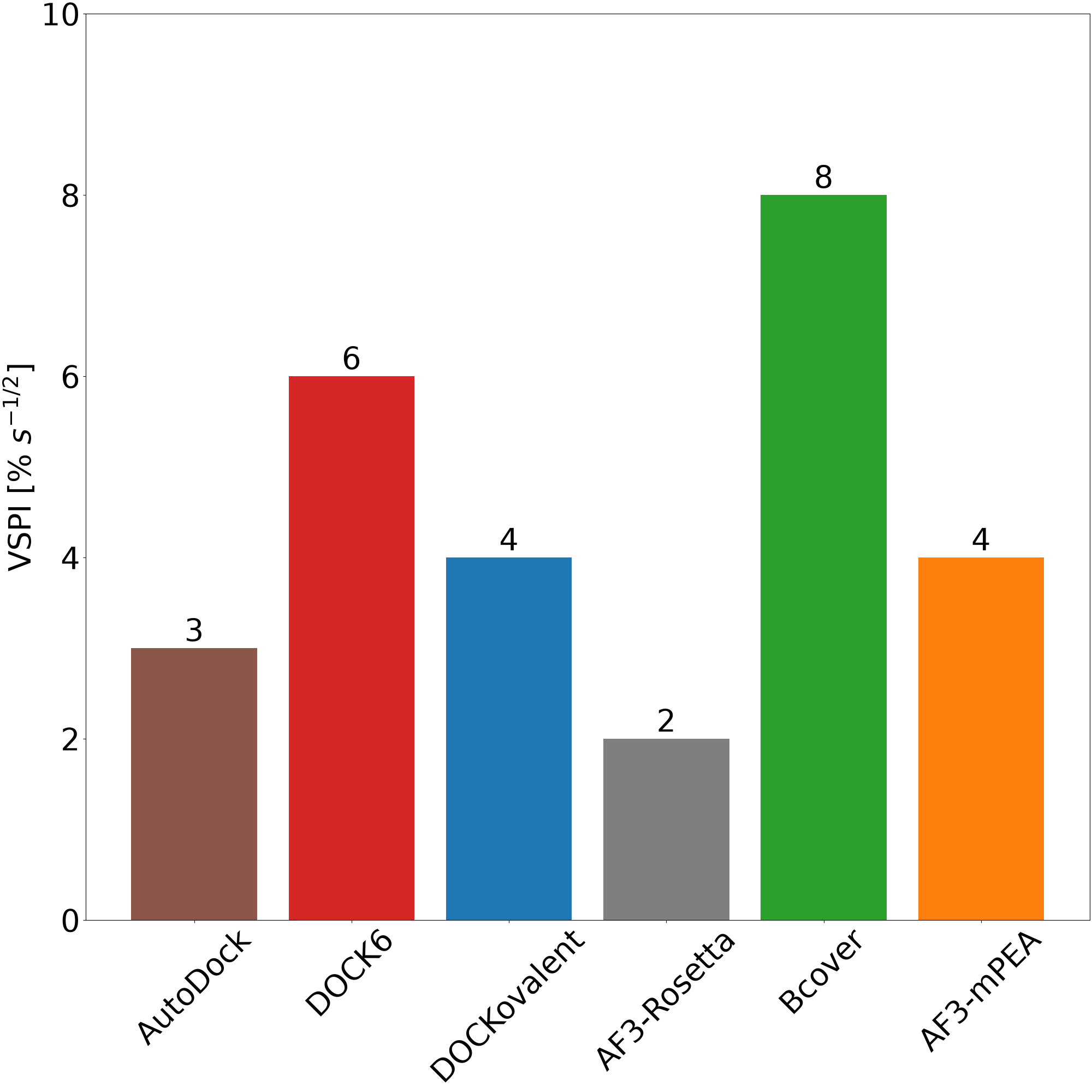}
    \caption{
    Average virtual-screening productivity index (VSPI) across COValid target sites.
    Bars compare the time-adjusted screening productivity of each method.}
    \label{fig:vspi_average}
\end{figure}

Per-target adjusted LogAUC values for \bcover{} and the comparator methods are shown in Fig.~\ref{fig:logauc_per_target}, while the corresponding average values across the COValid benchmark are summarized in Fig.~\ref{fig:logauc_average}. An overall comparison of the evaluated methods, including virtual-screening accuracy and computational cost, is presented in Supplementary information.
To jointly assess screening performance and computational efficiency, we introduce the \emph{virtual-screening productivity index} (VSPI), defined as $
\mathrm{VSPI} = \frac{\mathrm{LogAUC}}{\tau^{1/2}}$, where $\tau = \frac{t}{t_0}$
is the normalized computational cost per ligand, $t$ is the average runtime expressed in CPU equivalents on a standard 32-core processor, and $t_0 = 60~\mathrm{s}$ is the reference execution time. The square-root dependence quantifies the influence of runtime while still favoring computationally efficient methods, reflecting the practical observation that reducing execution time from minutes to seconds is generally more valuable than achieving an equivalent proportional speedup for already fast methods.
Per-target VSPI values are presented in Fig.~\ref{fig:vspi_all}, whereas the average VSPI over the complete COValid benchmark is shown in Fig.~\ref{fig:vspi_average}. In evaluating execution time BCover's performance was tested on possibly reproducing computing resources conditions as reported in ~\cite{Shamir2026COValid}.

\begin{table}[htbp]
\centering
\caption{Average retrospective screening performance across the nine COValid targets. Runtime is reported as average time per ligand.}
\label{tab:covalid_average_metrics}
\begin{tabular}{lrr}
\toprule
\textbf{Method} & \textbf{logAUC (\%)} & \textbf{Time per ligand (s)} \\
\midrule
Rowan~\cite{rowan_scientific_2026} & -6.0 & 60.0 \\
DOCKovalent & 12.0 & 8.4 \\
DOCK6 & 16.8 & 7.0 \\
AutoDock & 7.0 & 6.7 \\
AF3-mPAE & 72.0 & 280.0 \\
AF3-Rosetta & 28.0 & 280.0 \\
\bcover{} & 31.0 & 10.2 \\
\bottomrule
\end{tabular}
\end{table}

\begin{table}[htbp]
\centering
\caption{Performance on the best-performing COValid target for \bcover{}: KRasG12C. Runtime is reported as average time per ligand.}
\label{tab:covalid_kras_metrics}
\begin{tabular}{lrr}
\toprule
\textbf{Method} & \textbf{logAUC (\%)} & \textbf{Time per ligand (s)} \\
\midrule
DOCKovalent & 0.0 & 8.4 \\
DOCK6 & 0.0 & 7.0 \\
AutoDock & 9.0 & 6.7 \\
AF3-mPAE & 74.0 & 280.0 \\
AF3--Rosetta & 30.0 & 280.0 \\
\bcover{} & 59.0 & 10.2 \\
\bottomrule
\end{tabular}
\end{table}

\begin{table}[htbp]
\centering
\caption{Performance on the hardest COValid target for \bcover{}: FGFR4 Cys552. Runtime is reported as average time per ligand.}
\label{tab:covalid_fgfr4_c552_metrics}
\begin{tabular}{lrr}
\toprule
\textbf{Method} & \textbf{logAUC (\%)} & \textbf{Time per ligand (s)} \\
\midrule
DOCKovalent & 7.0 & 8.4 \\
DOCK6 & 52.9 & 7.0 \\
AutoDock & -3.8 & 6.7 \\
AF3-mPAE & 82.2 & 280.0 \\
AF3--Rosetta & 47.1 & 280.0 \\
\bcover{} & 13.1 & 10.2 \\
\bottomrule
\end{tabular}
\end{table}

\begin{table}[htbp]
\centering
\caption{Summary of \bcover{} performance metrics across COValid.}


\label{tab:bcover_summary_metrics}
\begin{tabular}{lr}
\toprule
\textbf{Metric} & \textbf{Value} \\
\midrule
Average AUC & 0.88 \\
Maximum AUC & 0.96 \\
Average logAUC (\%) & 31.0 \\
Maximum logAUC (\%) & 57.0 \\
Average EF1 & 18.0 \\
Maximum EF1 & 33.0 \\
Average BEDROC & 0.39 \\
Maximum BEDROC & 0.83 \\
Time per ligand (s) & 10.2 \\
\bottomrule
\end{tabular}
\end{table}

The benchmark results indicate that \bcover{} provides a favorable balance between enrichment quality and computational cost. Across the nine COValid targets, \bcover{} achieves an average logAUC of 31\%, outperforming other physics-based docking tools DOCKovalent, DOCK6, AutoDock, and Rowan, while remaining much faster than AlphaFold3-based methods. Although AF3-mPAE gives the strongest average enrichment, it requires substantially longer runtime per ligand; in contrast, \bcover{} operates in the same practical runtime regime as lightweight docking tools while delivering markedly stronger ranking performance. This tradeoff is captured by the VSPI analysis (cf. Fig~\ref{fig:vspi_all,fig:vspi_average}, where \bcover{} shows one of the strongest time-adjusted screening profiles among the tested methods. Explicit values for logAUC and execution times are compared in Table~\ref{tab:covalid_average_metrics}.

The target-level results further illustrate the practical value of \bcover{}. On KRAS, shown in Table~\ref{tab:covalid_kras_metrics}, \bcover{} reaches a logAUC of 59\%, substantially exceeding AF3-Rosetta~\cite{abramson2024alphafold3,davis2009rosettaligand} and all other docking packages, and approaching the enrichment obtained by AF3-mPAE at a fraction of the computational cost. The harder FGFR4 Cys552 case highlights that performance remains target dependent, but even there \bcover{} provides a useful enrichment tool, consistently outpeforming open source, Rowan, Carsi-dock, AutoDock~\cite{morris2009AutoDock4} and DOCKovalent. Lower logAUC values are a  diagnostic for identifying target classes or pocket geometries where additional calibration may be needed. 

In Table~\ref{tab:bcover_summary_metrics} we summarize the \bcover{}-only metrics, which show an encouraging average AUC of 0.89, maximum AUC of 0.96, average EF1 of 18, maximum EF1 of 36, and average BEDROC of 0.39, indicating that the method is not only separating actives from decoys globally, but also enriching active compounds early in the ranked list. Overall, these results support \bcover{} as a scorer that improves over conventional covalent docking scores while retaining throughput suitable for prospective virtual screening.

\subsection{Pose Quality}
\label{subsec:pose_quality}

Although the primary objective of \bcover{} is ligand ranking for covalent virtual screening, pose quality provides an important complementary diagnostic. A highly ranked ligand should not only receive a favorable score, but should also adopt a geometrically plausible binding mode in the vicinity of the reactive residue. We therefore evaluated the top-ranked redocked poses using ligand heavy-atom RMSD. This analysis complements the adjusted LogAUC, BEDROC, EF1, and ROC-AUC metrics, because successful enrichment does not necessarily imply accurate recovery of the experimental binding geometry.

\begin{figure}[htbp]
  \centering
  \includegraphics[width=0.45\linewidth]{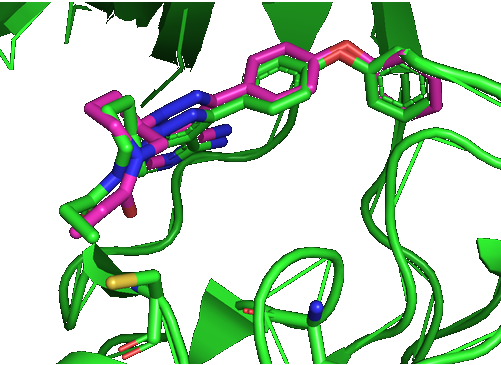}
  \caption{Redocking comparison for target 5P9J. The co-crystallized ligand pose is shown in green, and the redocked pose is shown in purple. The overlay illustrates the agreement between the experimentally observed binding geometry and the pose recovered by the docking protocol.}
  \label{fig:5p9j-redocking}
\end{figure}

For \bcover{}, redocking was completed successfully for nine of the ten investigated reference complexes; preprocessing failed for PDB entry 6MZW, which was therefore excluded from the RMSD analysis. Pose accuracy was quantified using two related measures. First, the ligand heavy-atom RMSD was calculated after optimal superposition of the predicted and crystallographic ligand conformations. Second, an unaligned RMSD was calculated in the receptor coordinate frame, thereby retaining differences in ligand translation and orientation within the binding pocket.

After optimal ligand alignment, \bcover{} achieved a mean RMSD of 1.28~\AA{} and a median RMSD of 1.46~\AA{} (Table~\ref{tab:pose_quality}). All nine successfully processed complexes had aligned RMSD values below 2.0~\AA{}, and three of nine complexes (33.3\%) were below 1.0~\AA{}. The corresponding unaligned RMSDs were higher, with a mean of 3.71~\AA{} and a median of 3.27~\AA{}. Two of nine poses (22.2\%) had unaligned RMSD values below 2.0~\AA{}, and one pose (11.1\%) was below 1.0~\AA{}.

The COValid study focused primarily on virtual-screening enrichment rather than on systematic redocking across all comparator methods, but it reported several useful pose-quality reference values~\cite{Shamir2026COValid}. In their study AlphaFold3 achieved a 78.5\% success rate on a covalent-complex benchmark when a successful prediction was defined as a pocket-aligned ligand RMSD below 2~\AA, following the AlphaFold3 benchmark protocol~\cite{abramson2024alphafold3,Shamir2026COValid}. Selected COValid examples included a pocket-aligned RMSD of 0.45~\AA{} for the KRAS G12C complex PDB 7O70, 0.67~\AA{} for an EGFR complex, and 0.84-1.33~\AA{} for MAP3K7 covalent complexes~\cite{Shamir2026COValid}.

These results indicate that \bcover{} consistently recovered ligand conformations similar to those observed crystallographically, whereas the absolute translation and orientation of the ligand within the binding pocket were less accurate for several systems. The strongest pose recovery was obtained for PDB entries 5P9J and 5JH6, with unaligned RMSDs of 0.61 and 1.19~\AA{}, respectively. By contrast, larger discrepancies between aligned and unaligned RMSDs for 5HG5, 6IUO, 6NVG, 8B78, and 8X2A indicate that the dominant error in these cases arose from ligand placement within the pocket rather than from the internal ligand conformation. Pose quality could be improved by refining the confining function for the NAC geometry constraint (cf. Fig.~\ref{fig:5p9j-redocking}).

Overall, the results support the use of \bcover{} for generating near-native ligand conformations with a reasonable quality, yet lower than AF3, marking the pre-reactive docking stage for further improvement.

\begin{table}[htbp]
\centering
\caption{Redocking pose accuracy of \bcover{} on the COValid reference structures. 
The aligned RMSD was calculated after optimal ligand superposition, whereas the 
unaligned RMSD retains the ligand translation and orientation in the receptor 
coordinate frame. RMSDs were calculated over ligand heavy atoms. PDB entry 6MZW 
was excluded because preprocessing did not complete successfully.}
\label{tab:pose_quality}
\begin{tabular}{lcc}
\toprule
\textbf{PDB entry} &
\textbf{Aligned RMSD (\AA)} &
\textbf{Unaligned RMSD (\AA)} \\
\midrule
4HCU & 1.69 & 2.39 \\
4QPS & 1.63 & 3.18 \\
5HG5 & 1.46 & 5.31 \\
5JH6 & 0.72 & 1.19 \\
5P9J & 0.61 & 0.61 \\
6IUO & 1.15 & 5.35 \\
6NVG & 1.94 & 4.90 \\
8B78 & 0.85 & 3.27 \\
8X2A & 1.46 & 7.16 \\
\midrule
\textbf{Mean} & \textbf{1.28} & \textbf{3.71} \\
\textbf{Median} & \textbf{1.46} & \textbf{3.27} \\
\bottomrule
\end{tabular}
\end{table}

Compared with classical covalent docking tools such as AutoDock, DOCK6, and DOCKovalent, \bcover{} is designed to make pose quality and reaction geometry explicit components of the scoring function rather than treating pose generation and ligand ranking as separable steps~\cite{morris2009AutoDock4,Allen2015DOCK6,London2014DOCKovalent}. This distinction is important because COValid demonstrates that strong enrichment and accurate pose recapitulation can be separate objectives~\cite{Shamir2026COValid}. A method can rank actives well while still producing local covalent geometries that require correction, and conversely, a near-native pose does not guarantee useful early enrichment. 

\subsection{Ablation of Scoring Terms}
We carried out an ablation study to quantify the contribution of individual components of the \bcover{} scoring model. In particular, we compared the full \bcover{} score against several reduced variants: pre-reactive docking-only, full score without Fukui-function terms, no-molecular-potential, and no-entropy models. Our objective was to test whether the Fukui-function and related reaction-aware terms improve ranking beyond docking and reactive geometry alone.

The pre-reactive docking-only variant uses only the quality of the non-covalent near-attack complex, without any explicit reactivity correction. The geometry-only variant retains local near-attack descriptors, such as the warhead-nucleophile distance and approach geometry, but removes electronic and thermodynamic terms. The no-Fukui and no-molecular-potential variants selectively remove the terms describing, respectively, the electrophilic reactivity of the warhead and the electrostatic organization of the protein pocket. Finally, the no-entropy variant removes the correction associated with conformational preorganization and hindered flexibility of the reactive complex.

The ablation results show that no single geometric descriptor is sufficient to recover the performance of the full model. Docking-only and geometry-only variants retain some early enrichment, confirming that formation of a plausible pre-reactive complex is a necessary condition for productive covalent binding. However, their performance is consistently lower than that of the full \bcover{} model, indicating that pose quality alone does not fully determine covalent reactivity. This is expected for covalent virtual screening: a ligand may adopt a geometrically plausible near-attack pose while still being electronically poorly activated, incorrectly polarized by the pocket, or entropically disfavoured.

The largest degradation was observed upon removing the entropic contributions, later the Fukui-function and molecular-potential terms. Removing the Fukui-function contribution alone reduced the LogAUC from $31.5$ to $28.3$. This indicates that the model benefits from explicitly resolving the electronic compatibility between the warhead and the reactive residue. The Fukui-function term captures the intrinsic local susceptibility of the ligand to nucleophilic attack, whereas the molecular-potential term captures how the protein environment stabilizes, polarizes, or destabilizes the pre-reactive charge distribution. Their removal therefore eliminates two complementary sources of information: ligand-local reactivity and pocket-induced electrostatic activation.
Removing the entropy term produced a 22.6 \% decrease in performance. This suggests that conformational preorganization significantly contributes to ranking too.

Overall, the ablation study supports the central design of \bcover{}: accurate covalent ranking requires combining three distinct layers of information. First, docking and near-attack geometry identify ligands capable of forming a plausible pre-reactive complex. Second, electronic descriptors, especially Fukui-function and pocket electrostatic-potential terms, identify complexes in which the warhead is chemically activated toward covalent bond formation. Third, entropy-related terms penalize poses that require excessive preorganization. The full model outperforms all reduced variants.

\section{Discussion}
\bcover{} was designed as a practical scoring model for covalent virtual screening, with the goal of improving early enrichment beyond docking-only or geometry-only ranking approaches~\cite{yu2019kinetics,mihalovits2020freeenergy,awoonorwilliams2021btk,voice2021ibrutinib}. The results of this study indicate that improved ranking requires the simultaneous inclusion of three complementary components: a warhead-reactivity descriptor, pocket-specific electrostatic information, and pre-reactive pose optimization. In the present implementation, warhead reactivity is represented by descriptors derived from the Fukui function~\cite{lonsdale2017warhead,palazzesi2019abinitio,hermann2020electrophilicity}, whereas the protein environment is described through the electrostatic potential of the binding pocket and an entropy model resolving torsional degrees of freedom. These physically interpretable quantities are combined in a regression-tree model, which provided the best overall performance among the tested scoring variants.

The strongest conclusion supported by our results is that electronic reactivity descriptors improve retrospective covalent-screening performance when they are used together with docking-derived pose information and local near-attack geometry constraints. The Fukui-function terms provide information about the intrinsic susceptibility of the warhead to nucleophilic attack, while the pocket electrostatic-potential terms describe how the protein environment modulates this reactivity. The entropy term provides an additional correction for conformational preorganization, helping to penalize poses that are geometrically plausible but unlikely to persist in a reactive configuration. Thus, \bcover{} should not be interpreted as a purely electronic reactivity model or as a conventional docking rescoring function. Rather, it combines geometric, electronic, and thermodynamic information into a single ranking for covalent ligand prioritization.

The present results suggest that \bcover{} is particularly well suited to kinase-focused covalent virtual screening, as supported by its performance on the Covalid Benchmark. The high enrichment observed for KRAS targets further suggests that the model can be useful in settings where productive covalent binding depends on a combination of precise near-attack geometry and warhead activation by the local pocket environment. At the same time, comparisons to experimental $K_i$, $k_{\mathrm{inact}}$, or $k_{\mathrm{inact}}/K_i$ values should be made only for datasets in which assay conditions, kinetic protocols, and compound annotations are sufficiently consistent and curated~\cite{mader2025assay,flanagan2014reactivegroups}. In its current form, \bcover{} is best viewed as a prioritization tool for virtual screening and hit triage. The highest-ranked compounds can then be advanced to more expensive and higher-fidelity calculations, including QM/MM and free-energy methods~\cite{evenseth2026covangelo}.

Several limitations should be noted. First, receptor flexibility and induced-fit effects are only approximately captured. Although docking and local geometry terms account for some aspects of pocket organization, large-scale conformational rearrangements, cryptic-pocket formation, and slow protein motions may require explicit ensemble or molecular-dynamics treatment. Second, protonation and tautomeric states can strongly affect computed reactivity descriptors and electrostatic potentials. These states must therefore be treated carefully during dataset preparation, especially for warheads and nearby catalytic residues. Third, the transferability of the model across warhead classes remains to be established. The present benchmark primarily supports conclusions for the reaction classes and target families represented in the benchmarked data; broader validation will be required before generalizing the model to substantially different electrophiles, nucleophiles, or reaction mechanisms.

Future work will focus on prospective experimental validation and on extending the benchmark to additional target families, broader kinetic datasets, and other warhead and reaction classes. We also plan to integrate \bcover{} with high-accuracy QM/MM free-energy methods~\cite{evenseth2026covangelo}, enabling a tiered strategy in which \bcover{} provides rapid enrichment during early screening, while more rigorous free-energy and reaction-barrier calculations are applied during lead optimization~\cite{chatterjee2017reversible,zhang2019reversible,mihalovits2020freeenergy}.

\section{Conclusions}
We have presented \bcover{}, a reaction-aware scoring tool for covalent virtual screening that combines classical docking terms with quantum-chemistry-derived reactivity descriptors. In contrast to conventional docking scores, the proposed method explicitly accounts for the interplay between electrophile reactivity, the local electrostatic environment of the protein pocket, and the geometry of the pre-reactive complex. The resulting scoring function remains computationally efficient while maintaining physically interpretable quantities directly related to covalent bond formation.

Retrospective evaluation on the COValid benchmark demonstrated that \bcover{} achieves competitive or superior enrichment compared with existing covalent docking approaches while maintaining throughput suitable for large-scale virtual screening. The modular design of the scoring function also allows straightforward extension to additional warhead types, reactive residues, and descriptor sets. We anticipate that \bcover{} will prove to be a practical tool for early-stage prioritization in covalent drug discovery, complementing more computationally demanding QM/MM and free-energy methods used in later-stage lead optimization.

\section{Acknowledgments}
We thank Jan Tulowiecki, Linn Evenseth and the rest of BEIT team for their support during the preparation of this manuscript.

\bibliography{references}

\end{document}